\documentclass[paper,11pt]{JHEP}
\usepackage{cite}
\usepackage[T1]{fontenc}
\usepackage[utf8]{inputenc}
\usepackage[english]{babel}
\usepackage[babel]{csquotes}
\usepackage{amsmath}
\usepackage{amsfonts}
\usepackage{amssymb}
\usepackage{mathtools}
\usepackage{mathrsfs}
\usepackage{bm}
\usepackage{bbold}
\usepackage{graphicx}
\usepackage{comment}
\usepackage{booktabs}
\usepackage{array}
\newcommand{\bra}{\langle}
\newcommand{\ket}{\rangle}

\newcommand{\varep}{\varepsilon}

\newcommand{\la}{\lambda}

\newcommand{\si}{\sigma}

\newcommand{\url}{\href}
\newcommand{\OO}{{\cal O}}
\newcommand{\pa}{\partial}

\newcommand{\beq}{\begin{equation}}
\newcommand{\eeq}{\end{equation}}

\newcommand{\mb}{\mathbb}
\newcommand{\mycomment}[1]{}
\newcommand{\wh}[1]{\widehat{#1}}

\title{Truncatable bootstrap equations in algebraic form  and critical surface 
 exponents}

\author{Ferdinando Gliozzi\\
 Dipartimento di Fisica, Universit\`a di Torino \\
and Istituto Nazionale di Fisica Nucleare - sezione di Torino\\  
Via P. Giuria 1 I-10125 Torino,Italy\\
e-mail:{\tt ferdinando.gliozzi@to.infn.it}}

\bigskip
\abstract{
We describe examples of drastic truncations of conformal bootstrap equations  encoding much more information than that obtained by a direct numerical approach. A three-term truncation of the four point function of a free  scalar  
in any space dimensions provides  algebraic identities among conformal block 
derivatives  which generate the exact spectrum of the infinitely many primary 
operators contributing to it. In  boundary conformal field theories, 
we point out that the appearance of free parameters in the solutions of bootstrap equations  
is not an artifact of truncations, rather it reflects a physical property of permeable conformal interfaces which are described by the same equations. Surface transitions correspond to isolated points in the parameter space. We are able to locate them in the case of 3d Ising model, thanks to a useful algebraic form of 3d boundary bootstrap equations.  
It turns out that the low-lying spectra of the surface operators in the 
ordinary and the special transitions of 3d Ising model form two different solutions of the same polynomial equation. 
Their interplay yields an estimate of the surface renormalization 
group exponents, $y_{h}=0.72558(18)$  for the ordinary universality 
class and $y_{h}=1.646(2)$ for the special universality class, which compare 
well with the most recent Monte Carlo calculations. 
Estimates of other surface exponents as well as OPE
coefficients are also obtained.}


\begin{document}

\section{Introduction}
\label{intro}
The precise computation of critical exponents in statistical mechanics or of anomalous dimensions in quantum field theory is in general a difficult task.
 With the exception of some two-dimensional models or some  supersymmetric theories, the best  results are traditionally obtained by epsilon-expansion techniques or 
Monte Carlo calculations. 

Assuming the theory under study  conformally invariant, one can in some  cases resort to bootstrap equations  \cite{Ferrara:1973yt} which provide constraints  on the possible conformal field theory (CFT) data, i.e.  the  spectrum of scaling dimensions of 
the local operators  contributing to a given $n$-point function and their 
operator product expansion (OPE) coefficients. A numerical approach to these 
equations first described in \cite{Rattazzi:2008pe}, relying on convex optimization, 
gave rise to a wealth of new  nontrivial and impressive  
results on CFTs in $d>2$ 
\cite{Rychkov:2009ij, Rattazzi:2010gj,Poland:2010wg,ElShowk:2012ht,Liendo
:2012hy,Pappadopulo:2012jk,ElShowk:2012hu, Gaiotto:2013nva, El-Showk:2014dwa,
Beem:2013qxa,Nakayama:2014yia,Nakayama:2014sba,Chester:2014fya,Kos:2014bka,Chester:2014gqa
,Beem:2014zpa,Simmons-Duffin:2015qma,Bobev:2015vsa,Kos:2015mba,Bobev:2015jxa,
Beem:2015aoa,Nakayama:2016jhq,Kos:2016ysd}. 
In particular its application to 3d Ising model, initiated in \cite{ElShowk:2012ht} and based on unitarity to find forbidden regions in the space of CFT data, 
led to very precise determinations of its bulk critical exponents \cite{Kos:2016ysd}.

There is another numerical approach \cite{Gliozzi:2013ysa,Gliozzi:2014jsa,Gliozzi:2015qsa} 
in which one takes into account the contribution to the bootstrap equations 
of just a handful  of low-dimension operators, compared to the hundreds of them contributing to the convex optimization, therefore it has been sometimes called 
``a severe truncation'' in the literature \cite{Rychkov:2015lca}. 
It yields less precise results in the case of  3d Ising model, moreover there is no systematic way of taking into account the error due to the truncation. 
However it applies also to non unitary theories. For instance the  scaling dimensions of low-lying local operators contributing to the 
Yang-Lee edge singularity in the whole range $2\le D\le 6$ have been 
evaluated  this way   
\cite{Gliozzi:2014jsa}. The results agree with strong coupling expansions and Monte Carlo simulations and were recently confirmed by  four loop calculations  
of $\phi^3$ field theory in six dimensions \cite{Gracey:2015tta}. Similarly, an evaluation  of the relevant   critical exponent of the ordinary surface transition 
of  3d Ising universality class has been obtained \cite{Gliozzi:2015qsa} in perfect agreement with the most recent Monte Carlo 
estimates \cite{Hasenbusch:2011o}. 
Good results are also reached by applying this numerical method to the 
study of the Yang-Lee model as well as  the critical Ising model on a three-dimensional projective space \cite{Nakayama:2016cim}.

Sometimes, though, the latter approach generates manifestly wrong 
results or, once added new terms to  truncation, free parameters arise in the solution, limiting  its predictive power.
In this paper we discuss some different instances of it, characterized 
 by the action of a hidden algebraic mechanism, each time different,
explaining such a behaviour. 

In the cases of free-field bosonic theory in $D$ dimensions or 2d CFTs , where surprising exact results go along some erroneous 
consequences, 
the hidden mechanisms 
consist in some unexpected algebraic identities among conformal blocks 
or their derivatives   which hold exactly at the symmetric point of the 
crossing symmetry. For instance, in the case of free-field theory, the 
identities built with a three-term truncation   
allow to obtain the exact spectrum of the whole infinite set of  
operators contributing to the scalar four-point function. 
However, the linear system employed for computing the OPE coefficients 
turns out to be ill-defined, so the standard numerical method to solve it 
gives erroneous results.      
Applying the bootstrap equations to another point alleviates this pathological behaviour, as explained in  Section 2.

The subsequent  Sections deal with boundary CFTs. Their truncated bootstrap equations are employed  to evaluate the spectrum of the low-lying surface operators in terms of the bulk spectrum, assumed to be known. At variance with the bulk case, one finds in general solutions depending on free parameters \cite{Gliozzi:2015qsa} which strongly limit their predictive power.

Here we want to emphasize that the appearance of free parameters is not an 
artifact generated by the truncation method, but rather reflects an unavoidable physical 
property of systems with conformal interfaces, i.e. scale 
invariant junctions of two CFTs, which are described by the same bootstrap equations. They are generalizations of a conformal boundary which corresponds to the case of a trivial theory in one side. 

Excitations propagating in the bulk are reflected by boundaries, while interfaces can be permeable, meaning that incident excitations are partly reflected and partly transmitted. The ``transmission coefficient'' \footnote{A definition of the transmission coefficient $T$ in terms of CFT data is available in 2d 
\cite{Quella:2006de}. Recently it was proven that this proposal obeys $T\le1$ for unitary theories \cite{Billo:2016cpy}.} generally 
depends on one or more parameters.  As a consequence, boundary conditions on interfaces are more general than those on a boundary, therefore the spectrum of surface operators is expected to be far richer than that on the boundary and to depend on free parameters. Examples of such interfaces have been discussed in condensed matter literature, see for instance \cite{Abe:1981,McCoy:1980ag,Oshikawa:1996dj}. 
 
In this paper we specialize on $D=3$ boundary CFTs, where we find 
 (Section 3) that the bootstrap equations can be written in a polynomial form, 
hence one can clearly see when and why the solutions of these 
equations depend on a free parameter. 
As a consequence, evaluating the relevant critical exponent of the ordinary transition in  $O(N)$ models is reduced 
to solving a simple quartic equation.

In Section 4, we point out that the surface exponents of the ordinary 
and special transitions of 3d Ising model are two different solutions of 
the same polynomial equation. Their interplay allows us to identify in the one-parameter family of solutions two points corresponding respectively to the 
special and to the ordinary transitions of 3d Ising model.  

Finally, in the last Section we summarize and compare our results with those obtained by field theoretical methods and Monte Carlo calculations.
We obtain in particular a precise evaluation of the leading magnetic 
exponents of ordinary and special transitions as well as 
the next to the leading universal corrections. Likewise, estimates of 
several OPE coefficients are also obtained.  The two magnetic exponents 
agree with those of most recent Monte Carlo estimates 
\cite{Hasenbusch:2011o,Hasenbusch:2011s}, while we were unable to find in 
literature other reliable estimates of the next to the leading exponents as 
well as of the  OPE coefficients.     
\section{ Exact truncations}
\label{exact}
In this Section we shall study some examples of the conformal four-point function of identical scalar operators, which can be parametrized as
\beq
\bra\phi(x_1)\phi(x_2)\phi(x_3)\phi(x_4)\ket=
\frac{g(u,v)}{ (x_{12}^2 x_{34}^2)^{\Delta_\phi}},
\label{fourpoint}
\eeq 
where $\Delta_\phi$ is the scaling  dimension of $\phi$,  $x_{ij}^2$ is the square of the distance between $x_i$ 
and $x_j$, $g(u,v)$ is a function of the cross-ratios 
$u=\frac{x_{12}^2x_{34}^2 }{x_{13}^2x_{24}^2 }$ and
 $v=\frac{x_{14}^2x_{23}^2 }{x_{13}^2x_{24}^2 }$.  
Using conformal invariance and OPE of $\phi(x)\phi(y)$ in the limit $x\to y$  one can derive the decomposition in conformal blocks $G_{\Delta,\ell}(u,v)$  i.e.
 eigenfunctions of the quadratic Casimir operator of $SO(D+1,1)$:
\beq
g(u,v)=1+\sum_{\Delta,\ell}{\sf p}_{\Delta,\ell}G_{\Delta,\ell}(u,v),
\label{bexpansion}
\eeq
with ${\sf p}_{\Delta,\ell}=\lambda_{\phi\phi\OO}^2$, where 
$\lambda_{\phi\phi\OO}$ is the coefficient of the primary operator 
$\OO=[\Delta,\ell]$ of scaling dimension $\Delta$ and spin $\ell$ contributing to the OPE of $\phi(x)\phi(y)$.

Invariance under permutations of the four $x_i$'s implies the following 
functional equations
\beq
g(u,v)=g(u/v,1/v)\,; \,\,v^{\Delta_\phi}g(u,v)=u^{\Delta_\phi}g(v,u)\,.
\label{fuequ}
\eeq
the first relation projects out the odd spins. 
The second one, after separating the identity from the other contributions, can be written as the sum rule
\beq
\sum_{\Delta,\ell}{\sf p}_{\Delta,\ell}
\frac{v^{\Delta_\phi}G_{\Delta,\ell}(u,v)-u^{\Delta_\phi}G_{\Delta,\ell}(v,u)}
{u^{\Delta_\phi}- v^{\Delta_\phi}}=1\,.
\label{sumrule}
\eeq
It is useful to adopt  the parametrization $u=z\bar{z}$ and 
$v=(1-z)(1-\bar{z})$ \cite{Dolan:2003hv} which simplifies the functional form 
of the conformal blocks. Following  \cite{ElShowk:2012ht} we make also the change of variables $z=(a+\sqrt{b})/2$, $\bar{z}=(a-\sqrt{b})/2$ and Taylor expand the sum rule around $a=1$ and $b=0$.
 It is easy to see that such an  expansion will contain only even powers 
of $(a-1)$ and integer powers of $b$.
The truncated sum rule can then be rewritten as a single inhomogeneous equation
\beq
\sum_{\Delta,\ell}{\sf p}_{\Delta,\ell}\,{\sf f}^{(0,0)}_{\Delta_\phi,\Delta_\ell}  =1,
\label{inho}
\eeq
which is employed to normalize the OPE coefficients,
and an infinite set of homogeneous equations
 \beq
{\sf f}\, {\sf p}\equiv\sum_{\Delta,\ell}{\sf f}^{(2m,n)}_{\Delta_\phi,\Delta_\ell}\,
{\sf p}_{\Delta,\ell} =0,~
(m,n\in\mb{N},m+n\not=0),
\label{homos}
\eeq
with
\beq
{\sf f}^{(m,n)}_{\alpha,\beta}=\left(\partial_a^{m}\partial_b^n
\frac{v^{\alpha}G_{\beta}(u,v)-u^{\alpha}G_{\beta}(v,u)}
{u^{\alpha}- v^{\alpha}}\right)_{a=1,b=0}.
\label{matrix}
\eeq
 According to the  numerical method described in \cite{Gliozzi:2013ysa} we truncate both the spectrum and the number of  homogeneous equations (\ref{homos}), 
keeping only the first $M$ 
derivatives and the first $N$ operators. Choosing $M\ge N$ the truncated homogeneous system admits a non-trivial solution only if all the minors of order $N$ vanish.  

In this Section we shall describe some exact truncations, i.e. a set of 
$M$ coinciding zeros of  minors of order $N$  
in which the spectrum $[\Delta_i,\ell_i]$, $(i=1,\dots,N)$ of the $N$ retained operators turn out 
to be identical with a subset of $N$ elements
 of the  infinite-dimensional, exact solution of the bootstrap equations.  

\begin{figure}
\centering
\includegraphics[width=10 cm]{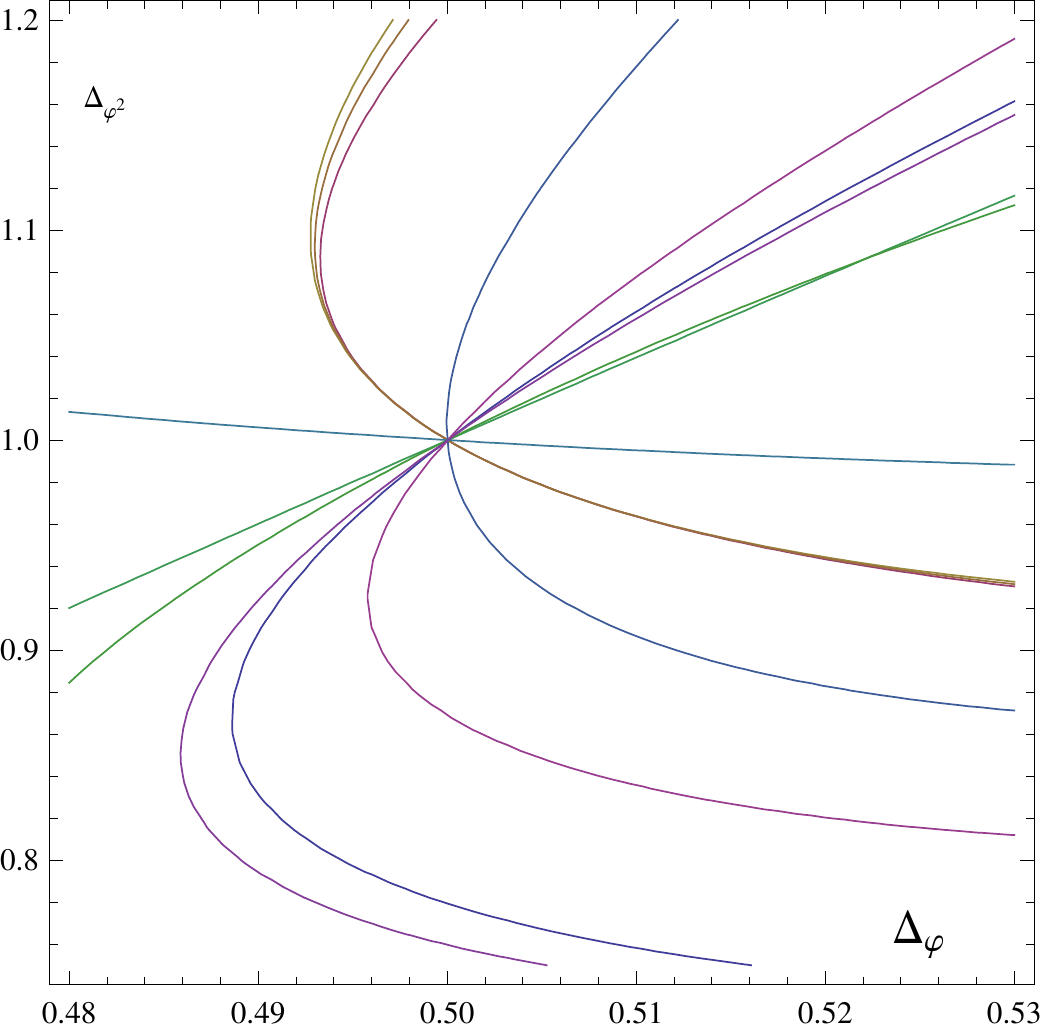}
\caption{Setting $D=3$ and $\Delta_4=\Delta_{\phi^2}+2$ in (\ref{freetruncation}) we get a two-dimensional section of the zeros of the 10 $3\times3$ minors  made with the first 5 derivatives specified in the text. The common intersection selects the exact solution of the free field theory.}
\label{fig:1}
\end{figure}

As a first example, consider a CFT  in $D$ space-time dimensions in which 
the scalar operator of Eq. (\ref{fourpoint}) has scaling dimension 
$\Delta=\frac{D-2}2$. It is known that this scalar is necessarily a free field (see for instance \cite{Weinberg:2012cd}), therefore using the explicit form of the free four-point function we can write the exact fusion rule
\beq
\left[\frac{D-2}2\right]\times\left[\frac{D-2}2\right]=
\sum_{k=0}^{\infty}[D-2+2k,\ell=2k]
\label{freed} 
\eeq
describing the set of primary operators $[\Delta,\ell]$  contributing to the OPE of
$\phi(x) \phi(y)$. On general grounds one would expect that a finite 
truncation of (\ref{freed})   perturbs someway the spectrum, 
therefore we write a $N=3$ truncation in the form
\beq
[\Delta_\phi]\times [\Delta_\phi]=[\Delta_{\phi^2}]+[D,2]+[\Delta_4,4]~,
\label{freetruncation}
\eeq 
where we only assumed the conservation of the energy-momentum tensor associated with the conformal block $[\Delta_2,2]$, which  entails $ \Delta_2=D$. 
This truncation depends on the three unknowns $\Delta_\phi, \Delta_{\phi^2}, \Delta_4$. We look for solutions of the first  $M=5$ homogeneous equations with $n+m\le2$. We can perform with them 10 different subsystems made with 3 equations and 3 unknowns. A plot of the different solutions is drawn in Figure \ref{fig:1}. Since there are more independent minors than unknowns we would expect a set of scattered solutions. 
Against all 
expectations, such a truncation admits a unique exact solution with  $\Delta_\phi=\frac{D-2}2,\Delta_{\phi^2}=D-2, \Delta_4=D+2$, like in the exact fusion rule 
(\ref{freed}). This can be easily verified by plugging in the homogeneous 
system (\ref{matrix}) the conformal blocks of the free theory in the cases
with integer $D$  where a closed form is known. In particular for $D=4$ and $D=6$  we used  the explicit formulae found in \cite{Dolan:2003hv} and in  
$D=3$ the closed formula found in \cite{Rychkov:2015lca} on the diagonal 
$z=\bar{z}=a/2$. In the latter case   a conformal block $g_\ell(z)\equiv 
G_{\ell+1,\ell}(u,v)\vert_{b=0}$ of spin $\ell$ and scaling dimension $\Delta=\ell+1$ becomes simply
\beq
 g_\ell(z)=\left(\frac{4z}{(1+\sqrt{1-z})^2}\right)^{\ell+1}
\frac{(1+\sqrt{1-z})^4}{(1+\sqrt{1-z})^4-z^2}.
\label{freeblock}
 \eeq  
We want now to manage the found solution in a fashion that  will allow us to reconstruct the whole spectrum contributing to the fusion rule (\ref{freed}) for any $D$.  The mentioned  10 different $3\times3$ subsystems (\ref{homos}) 
are associated with 
10 different matrices ${\sf f}_i$ $(i=1,\dots,10)$. At the  solution 
these ${\sf f}_i$'s admit a non-vanishing co-kernel or left null space, i.e.
  ${\sf v}_i{\sf f}_i=0$ (no sum on $i$ and ${\sf v}_i \not=0$). The left eigenvectors ${\sf v}_i$ span a 
3d subspace of the 5d linear space of  derivatives
${\sf f}^{(2m,n)}_{\alpha,\beta}\,, (m+n\le 2)$ defined in (\ref{matrix}). Choosing a basis, we obtain, for each conformal block contributing to the solution, three linearly independent relations

\begin{align}
&\frac12{\sf f}^{(2,0)}_{\Delta_\phi,\Delta_\ell}-
\frac{D-1}3\, {\sf f}^{(0,1)}_{\Delta_\phi,\Delta_\ell}=0\,;\label{identity1} \\
&\frac{D(4-D)}2\,{\sf f}^{(2,0)}_{\Delta_\phi,\Delta_\ell}-\frac58\,{\sf f}^{(4,0)}_{\Delta_\phi,\Delta_\ell} +
\frac{(D^2-1)}2\, {\sf f}^{(0,2)}_{\Delta_\phi,\Delta_\ell}=0\,;\label{identity2} \\
&3D\, {\sf f}^{(2,0)}_{\Delta_\phi,\Delta_\ell}+(D^2-1)\,{\sf f}^{(0,2)}_{\Delta_\phi,\Delta_\ell} -
\frac{3(D-1)}2\, {\sf f}^{(2,1)}_{\Delta_\phi,\Delta_\ell}=0\,.
\label{identity3}
\end{align}
Although by construction we would expect that they  are only identically satisfied  by the conformal blocks retained in the $N=3$ truncation, it turns out, unexpectedly, that the 
whole infinite set  of conformal blocks listed in (\ref{freed}) satisfies them. 
These surprising identities may be easily verified for the mentioned cases 
with integer $D$, as for instance (\ref{freeblock}). For arbitrary $D$ 
a straightforward check  may be obtained by directly applying the above 
relations  to the function  $g(u,v)$ defined  in 
(\ref{fourpoint}), which in a free field theory in $D$ dimensions is \[ g(u,v)=1+\left(\frac{u}v\right)^{\frac{D-2}2}+u^{\frac{D-2}2}.\]
It is also easy to  check that this solution does not depend on the specific form of the OPE coefficients by replacing  $g(u,v)$ with $P(C_2)\,g(u,v)$, where $P$ is an arbitrary polynomial of the Casimir operator $C_2$.

To sum up, we have described an example of  truncation that in some sense is 
exact seeing that it reproduces the true spectrum contributing to a four-point 
function.
There is however a price to be paid. The sum rule (\ref{sumrule}) tells us that
 truncating it to a finite number of terms  while keeping the true 
spectrum can not give the exact OPE coefficients. Actually the coefficients 
obtained solving the truncated equations for $D>3$ differ substantially from the exact values and the 
solution worsens as $D$ increases\footnote{ This problem was pointed out to the authors of  \cite{Gliozzi:2014jsa} by Yu Nakayama.}. Adding new  terms to the truncation does not help very much. In fact, the 
above identities imply  simple zeros for the $3\times3$ minors, double zeros for the  $4\times4$ minors and so on. In other terms 
the inhomogeneous system with $N>3$ depends on $N-3$ free parameters and loses part of its predictive power. It is worth noting that the above identities hold true only at the symmetric point $z=\bar{z}=\frac12$. In a different point it is no longer possible to reconstruct the true spectrum, the minors do not longer have multiple zeros and for  large enough truncations one may obtain reasonable values of the spectrum and  OPE couplings.   

As an aside, it is instructive to note that equations (\ref{identity1}),
(\ref{identity2}) and (\ref{identity3}) are analytic examples of linear functionals whose zeros define the exact spectrum of a whole sector of a CFT. They bear some similarity to the linear functional discussed in 
\cite{Rattazzi:2008pe,ElShowk:2012hu} in relation with the upper bound set by unitarity. Notice however that our functionals are not positive, with the only exception of eq.(\ref{identity1}) in $D=4$ which coincides, up to a suitable change of variables, with the one mentioned in  \cite{Rychkov:2009ij}, showing that the 4d free field  theory belongs to the unitarity upper bound.  

Another kind of pathological truncations may be found in 2d CFTs. Here it is known that the identity, the stress tensor as well as an infinite set of 
operators of known scaling dimension and spin are grouped into a single 
Virasoro conformal block, therefore  part of  terms contributing to 
the sum rule (\ref{sumrule}) is already known and one might envisage to apply the truncation method to get an approximate evaluation of the unknown part of the spectrum. It turns out that in a sufficiently large truncation all 
the minors of the homogeneous system
(\ref{homos}) made only with  $a$ derivatives vanish identically, irrespective of the scaling dimensions and spin of the unknown terms, so they do not provide any information about the unknown part of the spectrum. The explanation of this unexpected behaviour resides in an erratic identity among some terms of the mentioned Virasoro conformal block. Precisely, in the limit $b\to 0$ we have 
\beq
G_{4,0}(u,v)-\frac3{700} G_{8,0}(u,v)-\frac1{45} G_{6,2}(u,v)-G_{4,4}(u,v)+
\frac{25}{6237} G_{8,4}(u,v)=O(b).
\eeq
Thus all minors containing, besides other entries, only $a$ derivatives 
of the above  five conformal blocks are zero identically.  
In order to obtain useful information from the homogeneous 
system (\ref{homos}) we have to always include  $b$ derivatives.   

\section{Encoding bootstrap constraints into polynomial equations} 

The constraints imposed by conformal symmetry on correlation functions near a boundary were studied in \cite{McAvity:1995zd} and the boundary bootstrap program was set up in \cite{Liendo:2012hy}.  

The study of surface transitions of the 3d Ising and other $O(N)$ models through numerical solutions of truncated bootstrap equations \cite{Gliozzi:2015qsa} resulted particularly fruitful in the ordinary transition, where the knowledge of the low-lying bulk spectrum allowed to 
evaluate the scaling dimension of the relevant surface operator which compares well with known results of two-loop calculations \cite{Diehl:1998mh} and 
nicely agree with the more precise Monte Carlo estimates 
\cite{Hasenbusch:2011o}. 

In this Section we show 
that these equations may be written in a simple algebraic form, 
thereby enabling an analytic study of the solution.

A CFT in a semi-infinite $D$-dimensional space bounded by a flat 
$(D-1)$-dimensional  boundary is characterized by  the spectra of the bulk primary operators associated with  representations of the conformal group $SO(D+1,1)$ as well a that of the boundary primaries, associated with representations of $SO(D,1)$.
The two point function of  identical scalar primaries  can be parametrised as
\beq
\bra\phi(x)\,\phi(x')\ket=\frac{\xi^{-\Delta_\phi}}{(4 y y')^{\Delta_\phi}} g(\xi), 
\eeq    
where $\xi$ is the invariant combination 
\beq
\xi=\frac{(x-x')^2}{4 y y'}.
\eeq
$x$ denotes a point in the $D$ dimensional space  and $y$ its distance  
from the boundary.

\noindent $g(\xi)$ can be expanded either in terms of bulk conformal blocks or in terms of boundary blocks. More precisely we can write
\beq
  g(\xi)=1+\sum_\OO a_\OO\lambda_{\phi\phi\OO} f_{bulk}(\Delta_\OO,\xi),
\eeq
or
\beq
  g(\xi)=\xi^{\Delta_\phi}\left(a_\phi^2+\sum_{\widehat{\OO}} \mu_{\widehat{\OO}}^2 
f_{bdry}(\Delta_{\widehat{\OO}},\xi)\right),
\eeq
where we denoted the boundary quantities with a hat. 
Only scalar primaries contribute to both expansions. $\lambda_{\phi\phi\OO}$ 
is the OPE coefficient  already introduced in the four point 
function (\ref{fourpoint}) and similarly $\mu_{\widehat{\OO}}$ is the 
bulk-to-surface OPE coefficient; 
$a_\OO$ parametrises the one point function $\bra \OO(x)\ket=\frac{a_\OO}{(2y)^{\Delta_\OO}}$.  The conformal blocks  $f_{bulk}(\Delta,\xi)$ and 
$f_{bdry}(\Delta,\xi)$ are eigenfunctions of the bulk and the boundary quadratic Casimir operators, namely
\begin{align}
 {\cal C}_2\,f_{bulk}(\Delta,\xi)&=c_\Delta f_{bulk}(\Delta,\xi),~ c_\Delta=\Delta(D-\Delta),\\
 \wh{{\cal C}}_2\,f_{bdry}(\Delta,\xi)&=b_{\Delta} 
f_{bdry}(\Delta,\xi),~ b_\Delta=\Delta(D-1-\Delta).
\end{align}
These conformal blocks are completely fixed once  their asymptotic behaviour is given:
\beq
f_{bulk}(\Delta,\xi)\sim \xi^{\frac{\Delta}2}~(\xi\to 0);~~
f_{bdry}(\Delta,\xi)\sim \xi^{-\Delta}~(\xi\to \infty).
\eeq
Later we shall need the explicit form of  ${\cal C}_2$ and $\wh{\cal C}_2$.
 They can be easily obtained using the embedding formalism of  \cite{Dolan:2003hv} and \cite{Liendo:2012hy}
\begin{align}
{\cal C}_2&=-4\xi^2(1+\xi)\frac{d^2}{d\xi^2}-4\xi(\xi+1-D/2)
\frac{d}{d\xi}~;\label{bulkcasimir}\\
\wh{\cal C}_2&=-\xi(1+\xi)\frac{d^2}{d\xi^2}-D(\xi+\frac12)\frac{d}{d\xi}~.
\label{bdrycasimir}
\end{align}

The bootstrap constraint simply expresses the equality of the two above expansions.
It can be written again as a sum rule
\beq
\xi^{\Delta_\phi}\left(a_\phi^2+\sum_{\widehat{\OO}} \mu_{\widehat{\OO}}^2 
f_{bdry}(\Delta_{\widehat{\OO}},\xi)\right)-\sum_\OO a_\OO\lambda_{\phi\phi\OO} f_{bulk}(\Delta_\OO,\xi)=1.
\label{bsumrule}
\eeq
We want now apply such a constraint to the study of surface transitions in $D=3$.
First, we rewrite this functional equation in the  form of infinitely many 
linear equations, one for each coefficient of the Taylor expansion around, say, $\xi=1$.
Then, we truncate the two sums keeping only $n_{bulk}$ terms in the bulk 
channel and  $n_{bdry}$ terms in the boundary channel. We also truncate the Taylor expansion by keeping a finite number $M$ of derivatives.  We denote this 
truncation by the triple $(n_{bulk}, n_{bdry},s)$, where $s=1$ if the considered 
surface transition allows $a_\phi\not=0$, otherwise we set $s=0$. Using obvious shorthand notations, the truncated system may be written in the form
\beq
a^2_\phi+\sum_i^ {n_{bdry}}{\sf q}_i\,f_{bdry}(\Delta_i,\xi=1)-\sum_k^ {n_{bulk}}
{\sf p}_k\,
f_{bulk}(\Delta_k,\xi=1)=1,~~~~n_{bulk}+n_{bdry}+s=N
\label{binho}
\eeq
\beq
\pa_\xi^n\left[\xi^{\Delta_\phi}\left(a^2_\phi+\sum_i^ {n_{bdry}}{\sf q}_i\,f_{bdry}(\Delta_i,\xi)\right)-\sum_k^ {n_{bulk}}{\sf p}_k\,f_{bulk}(\Delta_k,\xi)\right]_{\vert_{\xi=1}}=0.
~~n=1.\dots,M
\label{bhomo}
\eeq
This truncation becomes predictive if one can find  solutions of the homogeneous system (\ref{bhomo}) with $M\ge N$. Interesting solutions have been found for 
$M=N+1$ or $M=N+2$, while truncations with $M>N+2$ do not yield reliable solutions \cite{Gliozzi:2015qsa}.

The simplest solution analyzed in  \cite{Gliozzi:2015qsa} is associated with the truncation $(2,1,0)$. It has been shown to give the scaling dimension of the relevant surface operator of the ordinary transition in terms of the scaling dimensions of $\phi,\epsilon$ and $\epsilon'$, where $\phi$ is the fundamental scalar,$\epsilon$ the energy operator, and $\epsilon'$ its first recurrence (related to the correction-to-scaling exponent $\omega$) of the 3d $O(N)$ theory.

We want to rewrite this solution in an algebraic form. First, note  that the Casimir operators (\ref{bulkcasimir}) and (\ref{bdrycasimir}) enable us to write the $n^{\rm th}$ derivative of the bulk or boundary conformal blocks as a linear combination of the conformal block and its first derivative
\beq
\frac{d^n}{d\xi^n}f_{bulk}={\sf P}_n(c)f_{bulk}+{\sf Q}_{n-1}(c)f_{bulk}';~
\frac{d^n}{d\xi^n}f_{bdry}=\wh{\sf P}_n(b)f_{bdry}+\wh{\sf Q}_{n-1}(b)f_{bdry}',
\label{poly}
\eeq
where ${\sf P}_n,{\sf Q}_n,\wh{\sf P}_n$ and  $\wh{\sf Q}_n$ are polynomials of degree $[\frac{n}2]$ on their argument. 

In $D=3$ $f_{bdry}(\Delta,\xi)$ can be expressed as elementary algebraic functions  \cite{Gliozzi:2015qsa}, namely
\beq
f_{bdry}(\Delta,\xi)=\frac1{2\sqrt{\xi}}\left(\frac4{1+\xi}\right)^{\Delta-\frac12}
\left[1+\sqrt{\frac\xi{1+\xi}}\right]^{-2(\Delta-1)}.
\eeq
It will entail a dramatic simplification of the bootstrap equations. 
We only need to take advantage of the following two identities
\beq
z\equiv\frac{f_{bdry}'(\Delta,\xi)\vert_{\xi\to 1}}{f_{bdry} (\Delta,1)}=-
\frac{3-2(1-\Delta)\sqrt{2}}4;~b\equiv(2-\Delta)\Delta=-\frac18-3z-2z^2,
\label{bide}
\eeq
to express $\frac{d^n}{d\xi^n}f_{bdry}$ as a polynomial in $\Delta$ 
(or in $z$). As a consequence, the homogeneous system (\ref{bhomo}) 
associated  with the (2,1,0) truncation can be written as
\beq
x\,{\sf p}_\epsilon {\sf Q}_{n-1}(c_\epsilon)+ y\,{\sf p}_{\epsilon'} {\sf Q}_{n-1}(c_{\epsilon'})+
{\sf p}_\epsilon {\sf P}_{n}(c_\epsilon)+
{\sf p}_{\epsilon'} {\sf P}_{n}(c_{\epsilon'})={\sf q}
{\cal P}_n(\Delta_\phi,\Delta),
\label{xy}
\eeq
 where ${\cal P}_n$ is a polynomial of degree $n$ in its arguments 
and we set
\beq
x\equiv\frac{f_{bulk}'(\Delta_\epsilon,\xi)\vert_{\xi\to1}}
{f_{bulk} (\Delta_\epsilon,1)},\,
y\equiv\frac{f_{bulk}'(\Delta_{\epsilon'},\xi)\vert_{\xi\to1}}
{f_{bulk} (\Delta_{\epsilon'},1)}.
\label{xydef}
\eeq
We can even eliminate 
$x$ and $y$  in (\ref{xy}) by combining the derivatives in such a way to form a 
power of ${\cal C}_2$, so to have
\beq
{\sf p}_\epsilon\,c_\epsilon^m+{\sf p}_{\epsilon'}\,c_{\epsilon'}^m={\sf q}\,{\cal Q}_{2m}(\Delta_\phi,\Delta),
\label{cc}
\eeq  
where  ${\cal Q}_{2m}$ is another polynomial of degree $2m$ in its arguments.

In order to find a solution of the (2,1,0) truncation  we have to choose three equations among the two sets (\ref{xy}) and (\ref{cc}), with the constraint that the order of the maximal derivative acting on $f_{bulk}$ and $f_{bdry}$ should not exceed $n_{bulk}+n_{bdry}+s+2=5$, therefore one equation, at least, should be of the type (\ref{xy}). We pick two equations of type (\ref{cc}) with $m=1,2$ and 
one of type (\ref{xy}) with $n<5$. In this way we can form four different 
subsystems of three equations in three unknowns. Note that the associated 
minors  are linear in $x$ and $y$. Taking any two of them we can solve for $x$ and $y$. Remarkably, the solution does not depend on the pair 
chosen. We have  
\beq
\frac2{c_\epsilon}x\equiv\frac{2 f_{bulk}'(\Delta_\epsilon,\xi)\vert_{\xi\to1}}{c_\epsilon\,f_{bulk} 
(\Delta_{\epsilon},1)}
=\frac{{\sf A}(c_{\epsilon'},
\Delta_\phi,z)}{{\sf B}(c_{\epsilon'},
\Delta_\phi,z)}\,,
\label{x}
\eeq
with
\beq
{\sf A}\left(c,\frac f4,z\right)=4-(3+2c)f+f^2-f^3-
(1+2c+f+3 f^2)4z-(2+3f)(4z)^2-(4z)^3
\eeq
and
\begin{align}
{\sf B}\left(c,\frac f4,z\right)=&21+2c -6(3+c)f+(17+2c)f^2-2f^3+f^4-16(1+c-(6+c)f-f^3)z\nonumber\\
+&2(5+c+3f+3f^2)(4z)^2+4(1+f)(4z)^3+(4z)^4.
\end{align}
Clearly the above relation defines an algebraic equation of degree four in 
$\Delta$, hence it can be solved exactly. Replacing the bulk quantities $\Delta_\phi,\Delta_\epsilon,\Delta_{\epsilon'}$ with the known values of the 3d $O(N)$ models one gets at once an estimate of  the scaling dimension of the relevant surface operator of the ordinary transition, which is almost identical with that 
obtained  in \cite{Gliozzi:2015qsa} using  numerical means. 

Being Eq. (\ref{x}) a quartic equation, it has other three roots. In the 3d Ising model one is at 
$\Delta\simeq 3.08$ (it could have some to do with the extraordinary transition, 
where $\Delta=3$ exactly, however it requires $a_\phi\not=0$ and is described by the truncation (2,1,1)). The other two are  complex conjugate  
with a real part $\Re e (\Delta)\simeq 0.35$, close to the expected value 
of the leading critical exponent of the special transition. One has to add more terms to the truncation in order to describe a full-fledged  special transition.
 
Besides eq.(\ref{x}) there is another similar equation generated by the present analytical method. It suffices to exchange $\epsilon\leftrightarrow\epsilon'$ in (\ref{x}). The estimate of $\Delta$ obtained this way is slightly shifted. Again, in order to reduce the spread between these two estimates one may try to find more accurate solutions by adding new bulk and/or boundary terms to the truncation,  however this addition  transforms the isolated solution discussed in this section into a one-parameter family of solutions. This issue will be 
discussed in the next section.

\section{Ordinary versus special transition}

The algebraic method described in the previous Section does not apply to other viable truncations, as it requires in general too high derivatives. For example, the (4,1,1) truncation utilized in \cite{Gliozzi:2015qsa}  to describe the extraordinary transition would  require 10 derivatives at least. 
A similar conclusion applies to the one-parameter family of solutions of 
(3,3,0) or (4,3,0) 
studied in \cite{Gliozzi:2015qsa}, or of (4,4,0) analyzed here, in relation with the special transition. Nevertheless in the latter case the peculiar algebraic properties of $f_{bdry}$ 
will help us to clarify some features of this kind of truncation. Eventually,
 it  will lead  to a precise estimate of the  scaling dimensions of the relevant surface operators in both  special  and  ordinary transitions.

Notice that one-parameter solutions of truncated bootstrap equations were also encountered in the study of four-point functions  \cite{Gliozzi:2014jsa}, however in that case the choice of the low-lying bulk spectrum fixed uniquely the value of the free parameter, while in the case of boundary CFTs the knowledge of the bulk spectrum does no longer suffice to  fix the surface spectrum. 
\begin{figure}
\centering
\includegraphics[width=10 cm]{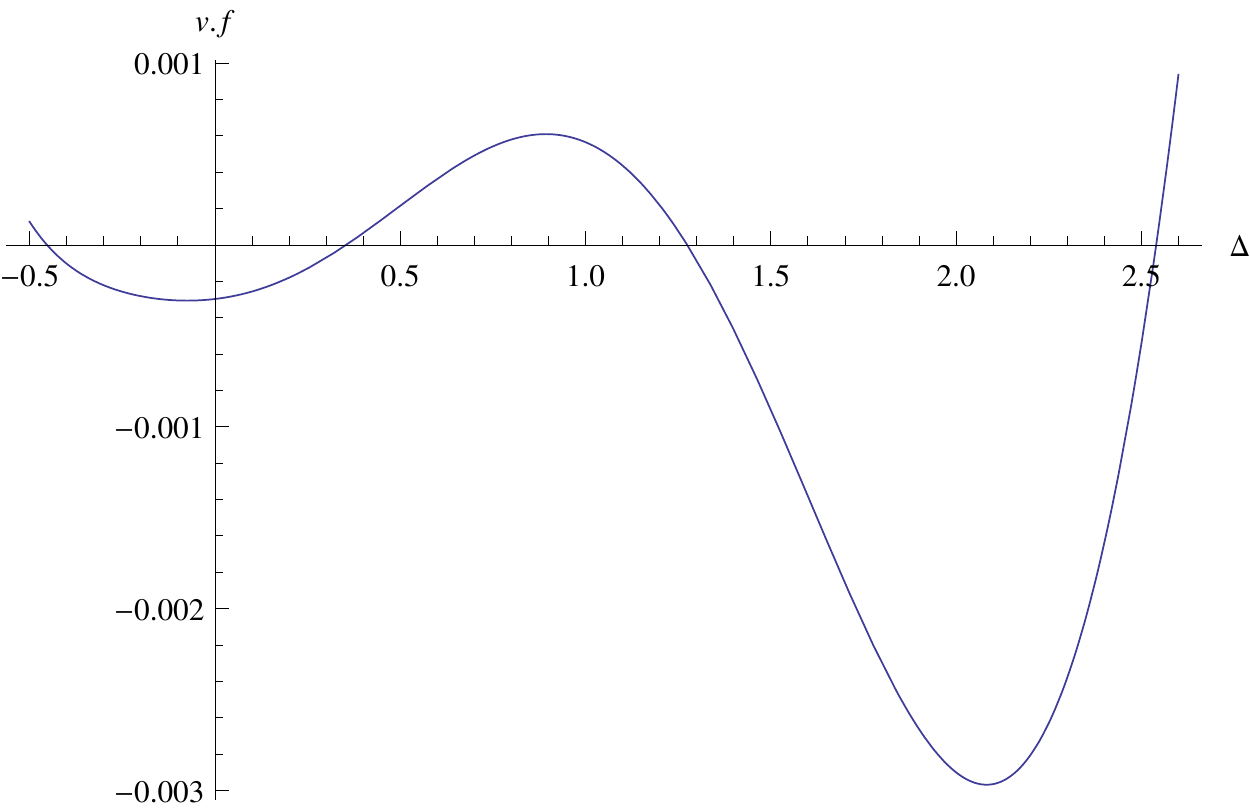}
\caption{The scalar product ${\sf v}\cdot f$ of a  null left eigenvector of the $8\times8$ 
minor ${\sf f}$ associated with the truncation (4,4,0) as a function of the 
scaling dimension of the surface operators.$f$ is the column vector defined in eq.(\ref{column}). Notice that 
${\sf v}\cdot { f}$   is a polynomial of $8^{\rm th}$ degree in $\Delta$, so it has more zeros than the four values that form a solution of (4,4,0). The additional zeros may be employed to form other solutions.}  
\label{fig:2}
\end{figure}

As already explained in the Introduction, a  physical motivation of such a  behaviour may be found in the fact  that the boundary bootstrap equations coincide with  interface bootstrap equations, 
where the conformal interface is 
the domain wall between the critical Ising model and another CFT. The main difference between a simple boundary or a more general  conformal interface of 3d Ising model resides in the boundary conditions, which are necessarily 
Dirichlet or Neumann in the former and more general in 
the latter. The case where  there is a free field theory in one side of 
the interface  has been analytically worked out in \cite{Gliozzi:2015qsa}. 
It turns out that its boundary conditions depend on a free parameter 
interpolating continuously between Dirichlet an Neumann conditions.   

Likewise the free parameter of the mentioned solutions is presumably a direct consequence of these more general boundary conditions. This is also supported by the fact that the solutions of the different truncations (3,3,0), (4,3,0) and (4,4,0) depend always on a single  parameter, though they deal with  different numbers of unknowns. Thus the problem of estimating the low-lying surface operators in the 
special transition can be reformulated, within this way of reasoning, in that of finding the value of the free parameter  corresponding to the Neumann conditions (the boundary conditions characterizing the special transition in the 
$\phi^4$ field theory).    

There is another element to be taken into account. Notice that the bootstrap equations of the special transition are the same of those 
 describing the ordinary transition. The solution is different, 
as the latter corresponds to Dirichlet boundary conditions. Therefore it 
should exist another value of the free parameter of the solution of, say, 
(4,4,0), describing also such a surface transition.
  
The numerical approach to solve truncated bootstrap equations is based on the Newton method, which requires as input a starting point with an 
approximate guess of the solution, thus it is not suitable for an exhaustive search. Here we describe a way out which takes advantage of the algebraic properties of the boundary conformal blocks.

In any CFT on a manifold with a boundary there is a short distance expansion expressing local bulk operators in terms of boundary operators. In the case of special transition of the critical 3d Ising model we have the fusion rule
\beq
\sigma\sim\wh\sigma+\wh\OO_s+\dots
\label{fusions}
\eeq  
where $\sigma$ is the lowest scalar in the bulk, $\wh\sigma$ the 
corresponding surface operator and $\wh\OO_s$ the next-to-leading surface primary.
Applying the direct numerical method of \cite{Gliozzi:2015qsa} to (3,4,0) 
or (4,4,0) truncations one may 
verify that unitary solutions exist only for  
$0.3\le\Delta_{\wh\sigma}\le0.45$  and  $2\le\Delta_{\wh\OO_s}\le2.6$. 
We have two other surface 
operators whose role is to provide more stable solutions. 

Every solution defines a left null eigenvector ${\sf v}$ of the $8\times8$ matrix ${\sf f}$ of the homogeneous system $(\ref{bhomo})$ such that ${\sf v\,f}=0$.  Define now the vector
\beq 
f (\Delta)=(\pa_\xi\,\,\xi^{\Delta_\sigma} f_{bdry},\pa^2_\xi\,\,\xi^{\Delta_\sigma} f_{bdry},
\dots  ,
\pa^8_\xi\,\,\xi^{\Delta_\sigma} f_{bdry}),
\label{column}
\eeq
which is a column of  ${\sf f}$.
 According to (\ref{poly}), the scalar product 
${\sf v\cdot}f$ is a polynomial of $8^{\rm th}$ degree  in $\Delta$, 
therefore it vanishes not only at the four points of the chosen solution, 
but also at other four values (some of them are plotted in Figure \ref{fig:2}). Actually every subset of four roots of this polynomial defines a solution. Most of them are uninteresting as they reveal some non-unitary features. There is 
however a solution  which is physically interesting. Its fusion rule in the boundary channel can be written as
\beq
\sigma\sim\frac{d{\wh\sigma}}{d y}+(\wh\OO_s)+\wh\OO_o+\dots
\label{fusiono}
\eeq
where the lowest surface operator is identified with the one of the ordinary transition since it has similar scaling dimensions and  similar OPE 
coefficients of the solution found in the previous Section. 

The term $\wh\OO_s$ coincides with that of (\ref{fusions}). We  put it within parentheses seeing that its square 
coupling $\mu^2_{\wh\OO_s}$ changes sign as we scan the family of solutions.

Actually the point where  $\mu^2_{\wh\OO_s}=0$, corresponding to the fusion
 rule $\sigma\sim\frac{d{\wh\sigma}}{d y}+\wh\OO_o+\dots$, selects the sought after 
solution, as it can be argued putting together the following three statements:
\begin{itemize}
\item[i)] According to the fusion rule (\ref{fusions}), $\wh\OO_s$ is non-vanishing on surfaces with Neumann boundary conditions, seeing that it is always present whenever $\wh\sigma$ is there.
\item[ii)] There is no reason to believe that an operator 
contributing to a surface transition with, say, Neumann boundary conditions could survive, with the same scaling dimension, on a surface with 
Dirichlet boundary conditions.
\item[iii)] The fusion rule (\ref{fusiono}) with $\mu^2_{\wh\OO_s}=0$ is the only solution compatible with  Dirichlet boundary conditions.  The corresponding value of $\Delta_{\OO_s}$ is the only one compatible with pure Neumann boundary conditions in (\ref{fusions}). 
\end{itemize}
In conclusion, we have selected in the one-parameter family for Ising 
conformal interfaces a pair of solutions 
describing the special and the ordinary transition. 
\section{Numerical results and concluding remarks}
In the previous Section we described a method to get  numerical estimates of the low-lying spectrum of surface operators in ordinary and special transitions  of 3d  Ising universality class in terms of the bulk spectrum. In table 1  we report the most significant quantities generated by the (4,4,0) truncation. 

 The input parameters are the scaling dimensions of the bulk operators 
$\sigma,\epsilon,\epsilon',\epsilon''$ and $\epsilon'''$, namely,
\[\Delta_\sigma=0.5181489(10);~\Delta_\epsilon=1.412625(10),\]
taken from  \cite{Kos:2016ysd}, $\Delta_\epsilon=3.8303(18)$ taken from 
\cite{El-Showk:2014dwa}, and $\Delta_{\epsilon''}=7.316(14);~
\Delta_{\epsilon'''}=13.05(4)$ from the truncation (4,1,1) studied in \cite{Gliozzi:2015qsa}.

\begin{table}[htb]
\centering

\begin{tabular}{*{5}{>{$}c<{$}}}
\toprule
\multicolumn{5}{c}{Ordinary surface transition in 3d Ising bulk universality class} \\
\midrule
\textup{Ref}  &   \textup{year}&\textup{method}& y_h=
2-\Delta_{\frac{d{\wh\sigma}}{d y}}&\Delta_{\OO_o}  \\
\textup{\cite{Diehl:1981jgg}} &    1981 & \varepsilon- \textup{expansion}& 0.762& \\
\textup{\cite{Diehl:1998mh}} &    1998 & \textup{3d two-loop exp.}         &   0.714&  \\
\textup{\cite{Deng:2005dh}} &    2005 & \textup{Monte Carlo}           &   0.7374(15)&  \\
\textup{\cite{Hasenbusch:2011o}}  &    2011 & \textup{Monte Carlo}           &    
0.7249(6)&  \\
\textup{\cite{Gliozzi:2015qsa}} &     2015 & \textup{Bootstrap}    
&   0.724(2)&\\
 \textup{this work}&      2016 & \textup{Bootstrap}    & 0.72558(18)&5.13(20)\\
\bottomrule
\end{tabular} \hspace{0.16\columnwidth}
\begin{tabular}{*{6}{>{$}c<{$}}}
\toprule
\multicolumn{6}{c}{Special surface transition in 3d Ising bulk universality class} \\
\midrule
\textup{Ref}  &   \textup{year}&\textup{method}& y_h=2-\Delta_{\wh\sigma}&y_t&\Delta_{\OO_s}  \\
 \textup{\cite{Diehl:1981,Diehl:1983}}  &    1983 & 
\varepsilon- \textup{expansion} &  1.65&1.08  & \\
\textup{\cite{Diehl:1998mh}} &    1998 & \textup{3d two-loop exp.} &   
1.583&0.856&\\
 \textup{\cite{Deng:2005dh}} &    2005 & \textup{Monte Carlo} &   
1.636(91)&0.715(1)& \\
\textup{\cite{Hasenbusch:2011s}} & 2011 & \textup{Monte Carlo}& 1.6465(6)&0.718(2)&\\
\textup{\cite{Gliozzi:2015qsa}} & 2015 & \textup{Bootstrap}& 1.55\div 1.7&&2\div2.6\\
 \textup{this work}&      2016 & \textup{Bootstrap}    & 1.6460(20)&&2.54(2)\\
\bottomrule
\end{tabular} \hspace{0.16\columnwidth}
\begin{tabular}{*{5}{>{$}c<{$}}}
\toprule
\textup{Surface transition}  & a_\varep\la_{\sigma\sigma\varep}  &   a_{\varep'} \la_{\si\si\varep'}  &
 \mu_1^2&\mu_2^2 \\
\midrule
\textup{Ordinary}  &   -0.7862(2)& 0.0413(1)& 0.7560(1) &0.000049(6)  \\
\textup{Special}  &  1.216(13) &  0.081(3)& 2.06(1)&    0.0251(7)    \\
\bottomrule
\end{tabular}
\caption{In the first two tables we compare our results with those obtained by field theoretic methods and Monte Carlo studies.  The last table collects the most relevant OPE coefficients. In the ordinary transition 
$\mu_1=\mu_{\frac{d\wh\sigma}{dy}}$ and $\mu_2=\mu_{\wh\OO_o}$, while in the special transition $\mu_1=\mu_{\wh\sigma}$ and $\mu_2=\mu_{\wh\OO_s}$. The first three OPE 
coefficients in the ordinary transition agree with those computed in \cite{Gliozzi:2015qsa}. }
\label{tab:1}
\end{table}

In the first two tables we compare our results for the relevant surface 
exponents of the two transitions with those obtained  with perturbative expansions in $\phi^4$ field theory or computed in Monte Carlo simulations. Following the literature, here we replaced the scaling dimension 
$\Delta$  with the surface renormalization 
group exponent $y_h=2-\Delta$. The field-theoretic approach to critical behaviour near free surfaces is rather challenging, mainly because the breakdown of translational invariance gives rise to serious computational difficulties. As a matter of fact expansions in 
4-$\varepsilon$  dimensions \cite{Diehl:1981jgg,Diehl:1981,Diehl:1983} or 3d expansions in a massive model\cite{Diehl:1998mh} do not exceed second order. Better results are obtained with Monte Carlo simulations 
\cite{Deng:2005dh,Hasenbusch:2011o,Hasenbusch:2011s}.

The error in our bootstrap results is due to the uncertainty of the input parameters. The major sources of error are the uncertainties of the scaling 
dimensions of $\epsilon''$ and $\epsilon'''$, while the errors and the central values do not vary
 sensibly when  we replace the values of $\Delta_\sigma$ 
and $\Delta_\epsilon$ with the estimates obtained in Monte Carlo 
simulations \cite{Hasenbusch:2011yya}. 

The method described in this work, being based on the interplay between 
ordinary and special transition, can not be directly generalized to $O(N)$ models with $N>1$. In fact the special transition is an order-disorder transition 
on the surface boundary, while there can not be any spontaneous breaking of a continuous symmetry in $D\le2$ dimensions, according to the Mermin-Wagner-Hohenberg theorem. Actually the special transition, being multicritical, is characterized by two relevant renormalization group exponents, the magnetic ($y_h$) and the thermal 
($y_t$) exponents. The latter is related to the fusion rule of the bulk energy 
in the boundary channel $ \epsilon\sim 1+\wh\epsilon+\dots$ by 
$y_t=2-\Delta_{\wh\epsilon}$. For completeness some values of $y_t$ are reported in the second table. So far, no suitable solution of the bootstrap equations  
has been found to compute  $\Delta_{\wh\epsilon}$. The algebraic formalism developed in this paper could help to face this problem.

\section*{Acknowledgements}
This work was initiated  at the workshop ``Bootstrap 2015'', May 18-29  2015 at the Weizmann Institute of Science in Rehovot, Israel, and preliminary results were first 
presented at the mini-conference on Statistical Physics at SISSA, Trieste, Italy, 9-10 October 2015. I 
thank the organizers and the participants of these two workshops for the stimulating atmosphere.  The final form has been presented at the GGI workshop on ``Conformal Field Theories and Renormalization Group Flows in Dimensions $d>2$\,'', Florence,  May 23- July 8 2016.
I would like to  thank the Galileo Galilei Institute for hospitality and the Simons Foundation for support during the completion of this work.

\providecommand{\href}[2]{#2}\begingroup\raggedright\endgroup


\begin{thebibliography}{10}

\bibitem{Ferrara:1973yt}
S.~Ferrara, A.~Grillo, and R.~Gatto, \emph{{Tensor representations of conformal
  algebra and conformally covariant operator product expansion}},
\href{http://dx.doi.org/10.1016/0003-4916(73)90446-6}{Annals Phys. {\bf 76}
  (1973)  161--188}.

\bibitem{Rattazzi:2008pe}
R.~Rattazzi, V.~S. Rychkov, E.~Tonni, and A.~Vichi, \emph{{Bounding scalar
  operator dimensions in 4D CFT}},
  \href{http://dx.doi.org/10.1088/1126-6708/2008/12/031}{JHEP {\bf 0812} (2008)
   031},
\href{http://arxiv.org/abs/0807.0004}{{\tt arXiv:0807.0004 [hep-th]}}.

\bibitem{Rychkov:2009ij}
V.~S. Rychkov and A.~Vichi, \emph{{Universal Constraints on Conformal Operator
  Dimensions}}, \href{http://dx.doi.org/10.1103/PhysRevD.80.045006}{Phys.Rev.
  {\bf D80} (2009)  045006},
\href{http://arxiv.org/abs/0905.2211}{{\tt arXiv:0905.2211 [hep-th]}}.

\bibitem{Rattazzi:2010gj}
R.~Rattazzi, S.~Rychkov, and A.~Vichi, \emph{{Central Charge Bounds in 4D
  Conformal Field Theory}},
  \href{http://dx.doi.org/10.1103/PhysRevD.83.046011}{Phys.Rev. {\bf D83}
  (2011)  046011},
\href{http://arxiv.org/abs/1009.2725}{{\tt arXiv:1009.2725 [hep-th]}}.

\bibitem{Poland:2010wg}
D.~Poland and D.~Simmons-Duffin, \emph{{Bounds on 4D Conformal and
  Superconformal Field Theories}},
  \href{http://dx.doi.org/10.1007/JHEP05(2011)017}{JHEP {\bf 1105} (2011)
  017},
\href{http://arxiv.org/abs/1009.2087}{{\tt arXiv:1009.2087 [hep-th]}}.

\bibitem{ElShowk:2012ht}
S.~El-Showk, M.~F. Paulos, D.~Poland, S.~Rychkov, D.~Simmons-Duffin, A. Vichi, \emph{{Solving the 3D Ising Model with the Conformal Bootstrap}},
  \href{http://dx.doi.org/10.1103/PhysRevD.86.025022}{Phys.Rev. {\bf D86}
  (2012)  025022},
\href{http://arxiv.org/abs/1203.6064}{{\tt arXiv:1203.6064 [hep-th]}}.

\bibitem{Liendo:2012hy}
P.~Liendo, L.~Rastelli, and B.~C. van Rees, \emph{{The Bootstrap Program for
  Boundary $CFT_d$}}, \href{http://dx.doi.org/10.1007/JHEP07(2013)113}{JHEP
  {\bf 1307} (2013)  113},
\href{http://arxiv.org/abs/1210.4258}{{\tt arXiv:1210.4258 [hep-th]}}.

\bibitem{Pappadopulo:2012jk}
D.~Pappadopulo, S.~Rychkov, J.~Espin, and R.~Rattazzi, \emph{{OPE Convergence
  in Conformal Field Theory}},
  \href{http://dx.doi.org/10.1103/PhysRevD.86.105043}{Phys.Rev. {\bf D86}
  (2012)  105043},
\href{http://arxiv.org/abs/1208.6449}{{\tt arXiv:1208.6449 [hep-th]}}.

\bibitem{ElShowk:2012hu}
S.~El-Showk and M.~F. Paulos, \emph{{Bootstrapping Conformal Field Theories
  with the Extremal Functional Method}},
  \href{http://dx.doi.org/10.1103/PhysRevLett.111.241601}{Phys.Rev.Lett. {\bf
  111} (2013) no.~24, 241601},
\href{http://arxiv.org/abs/1211.2810}{{\tt arXiv:1211.2810 [hep-th]}}.

\bibitem{Gaiotto:2013nva}
D.~Gaiotto, D.~Mazac, and M.~F. Paulos, \emph{{Bootstrapping the 3d Ising twist
  defect}}, \href{http://dx.doi.org/10.1007/JHEP03(2014)100}{JHEP {\bf 1403}
  (2014)  100},
\href{http://arxiv.org/abs/1310.5078}{{\tt arXiv:1310.5078 [hep-th]}}.

\bibitem{El-Showk:2014dwa}
S.~El-Showk, M.~F. Paulos, D.~Poland, S.~Rychkov, D.~Simmons-Duffin, and
  A.~Vichi, \emph{{Solving the 3d Ising Model with the Conformal Bootstrap II.
  c-Minimization and Precise Critical Exponents}},
  \href{http://dx.doi.org/10.1007/s10955-014-1042-7}{J. Stat. Phys. {\bf 157}
  (2014)  869},
\href{http://arxiv.org/abs/1403.4545}{{\tt arXiv:1403.4545 [hep-th]}}.

\bibitem{Beem:2013qxa}
C.~Beem, L.~Rastelli, and B.~C. van Rees, \emph{{The $\mathcal N=4$
  Superconformal Bootstrap}},
  \href{http://dx.doi.org/10.1103/PhysRevLett.111.071601}{Phys.Rev.Lett. {\bf
  111} (2013)  071601},
\href{http://arxiv.org/abs/1304.1803}{{\tt arXiv:1304.1803 [hep-th]}}.

\bibitem{Nakayama:2014yia}
Y.~Nakayama and T.~Ohtsuki, \emph{{Five dimensional $O(N)$-symmetric CFTs from
  conformal bootstrap}},
  \href{http://dx.doi.org/10.1016/j.physletb.2014.05.058}{Phys.Lett. {\bf B734}
  (2014)  193--197},
\href{http://arxiv.org/abs/1404.5201}{{\tt arXiv:1404.5201 [hep-th]}}.

\bibitem{Nakayama:2014sba}
Y.~Nakayama and T.~Ohtsuki, \emph{{Bootstrapping phase transitions in QCD and
  frustrated spin systems}},
  \href{http://dx.doi.org/10.1103/PhysRevD.91.021901}{Phys. Rev. {\bf D91}
  (2015) no.~2, 021901},
\href{http://arxiv.org/abs/1407.6195}{{\tt arXiv:1407.6195 [hep-th]}}.

\bibitem{Chester:2014fya}
S.~M. Chester, J.~Lee, S.~S. Pufu, and R.~Yacoby, \emph{{The $ \mathcal{N}=8 $
  superconformal bootstrap in three dimensions}},
  \href{http://dx.doi.org/10.1007/JHEP09(2014)143}{JHEP {\bf 1409} (2014)
  143},
\href{http://arxiv.org/abs/1406.4814}{{\tt arXiv:1406.4814 [hep-th]}}.

\bibitem{Kos:2014bka}
F.~Kos, D.~Poland, and D.~Simmons-Duffin, \emph{{Bootstrapping Mixed
  Correlators in the 3D Ising Model}},
  \href{http://dx.doi.org/10.1007/JHEP11(2014)109}{JHEP {\bf 11} (2014)  109},
\href{http://arxiv.org/abs/1406.4858}{{\tt arXiv:1406.4858 [hep-th]}}.

\bibitem{Chester:2014gqa}
S.~M. Chester, S.~S. Pufu, and R.~Yacoby, \emph{{Bootstrapping O(N) Vector
  Models in 4 < d < 6}},
\href{http://arxiv.org/abs/1412.7746}{{\tt arXiv:1412.7746 [hep-th]}}.

\bibitem{Beem:2014zpa}
C.~Beem, M.~Lemos, P.~Liendo, L.~Rastelli, and B.~C. van Rees, \emph{{The $
  \mathcal{N}=2 $ superconformal bootstrap}},
  \href{http://dx.doi.org/10.1007/JHEP03(2016)183}{JHEP {\bf 03} (2016)  183},
\href{http://arxiv.org/abs/1412.7541}{{\tt arXiv:1412.7541 [hep-th]}}.

\bibitem{Simmons-Duffin:2015qma}
D.~Simmons-Duffin, \emph{{A Semidefinite Program Solver for the Conformal
  Bootstrap}}, \href{http://dx.doi.org/10.1007/JHEP06(2015)174}{JHEP {\bf 06}
  (2015)  174},
\href{http://arxiv.org/abs/1502.02033}{{\tt arXiv:1502.02033 [hep-th]}}.

\bibitem{Bobev:2015vsa}
N.~Bobev, S.~El-Showk, D.~Mazac, and M.~F. Paulos, \emph{{Bootstrapping the
  Three-Dimensional Supersymmetric Ising Model}},
  \href{http://dx.doi.org/10.1103/PhysRevLett.115.051601}{Phys. Rev. Lett. {\bf
  115} (2015) no.~5, 051601},
\href{http://arxiv.org/abs/1502.04124}{{\tt arXiv:1502.04124 [hep-th]}}.

\bibitem{Kos:2015mba}
F.~Kos, D.~Poland, D.~Simmons-Duffin, and A.~Vichi, \emph{{Bootstrapping the
  O(N) Archipelago}}, \href{http://dx.doi.org/10.1007/JHEP11(2015)106}{JHEP
  {\bf 11} (2015)  106},
\href{http://arxiv.org/abs/1504.07997}{{\tt arXiv:1504.07997 [hep-th]}}.

\bibitem{Bobev:2015jxa}
N.~Bobev, S.~El-Showk, D.~Mazac, and M.~F. Paulos, \emph{{Bootstrapping SCFTs
  with Four Supercharges}},
  \href{http://dx.doi.org/10.1007/JHEP08(2015)142}{JHEP {\bf 08} (2015)  142},
\href{http://arxiv.org/abs/1503.02081}{{\tt arXiv:1503.02081 [hep-th]}}.

\bibitem{Beem:2015aoa}
C.~Beem, M.~Lemos, L.~Rastelli, and B.~C. van Rees, \emph{{The (2, 0)
  superconformal bootstrap}},
  \href{http://dx.doi.org/10.1103/PhysRevD.93.025016}{Phys. Rev. {\bf D93}
  (2016) no.~2, 025016},
\href{http://arxiv.org/abs/1507.05637}{{\tt arXiv:1507.05637 [hep-th]}}.

\bibitem{Nakayama:2016jhq}
Y.~Nakayama and T.~Ohtsuki, \emph{{Conformal Bootstrap Dashing Hopes of
  Emergent Symmetry}},
\href{http://arxiv.org/abs/1602.07295}{{\tt arXiv:1602.07295
  [cond-mat.str-el]}}.

\bibitem{Kos:2016ysd}
F.~Kos, D.~Poland, D.~Simmons-Duffin, and A.~Vichi, \emph{{Precision Islands in
  the Ising and $O(N)$ Models}},
\href{http://arxiv.org/abs/1603.04436}{{\tt arXiv:1603.04436 [hep-th]}}.

\bibitem{Gliozzi:2013ysa}
F.~Gliozzi, \emph{{More constraining conformal bootstrap}},
  \href{http://dx.doi.org/10.1103/PhysRevLett.111.161602}{Phys.Rev.Lett. {\bf
  111} (2013)  161602},
\href{http://arxiv.org/abs/1307.3111}{{\tt arXiv:1307.3111}}.

\bibitem{Gliozzi:2014jsa}
F.~Gliozzi and A.~Rago, \emph{{Critical exponents of the 3d Ising and related
  models from Conformal Bootstrap}},
  \href{http://dx.doi.org/10.1007/JHEP10(2014)042}{JHEP {\bf 1410} (2014)  42},
\href{http://arxiv.org/abs/1403.6003}{{\tt arXiv:1403.6003 [hep-th]}}.

\bibitem{Gliozzi:2015qsa}
F.~Gliozzi, P.~Liendo, M.~Meineri, and A.~Rago, \emph{{Boundary and Interface
  CFTs from the Conformal Bootstrap}},
  \href{http://dx.doi.org/10.1007/JHEP05(2015)036}{JHEP {\bf 05} (2015)  036},
\href{http://arxiv.org/abs/1502.07217}{{\tt arXiv:1502.07217 [hep-th]}}.

\bibitem{Rychkov:2015lca}
S.~Rychkov and P.~Yvernay, \emph{{Remarks on the Convergence Properties of the
  Conformal Block Expansion}},
  \href{http://dx.doi.org/10.1016/j.physletb.2016.01.004}{Phys. Lett. {\bf
  B753} (2016)  682--686},
\href{http://arxiv.org/abs/1510.08486}{{\tt arXiv:1510.08486 [hep-th]}}.

\bibitem{Gracey:2015tta}
J.~A. Gracey, \emph{{Four loop renormalization of $\phi^3$ theory in six
  dimensions}}, \href{http://dx.doi.org/10.1103/PhysRevD.92.025012}{Phys. Rev.
  {\bf D92} (2015) no.~2, 025012},
\href{http://arxiv.org/abs/1506.03357}{{\tt arXiv:1506.03357 [hep-th]}}.

\bibitem{Hasenbusch:2011o}
M.~Hasenbusch, \emph{{The  thermodynamic Casimir force: A Monte Carlo study of
  the crossover between the ordinary and the normal surface universality
  class}}, \href{http://dx.doi.org/10.1103/PhysRevB.83.134425}{Phys.Rev. {\bf B
  83} (2011)  134425},
\href{http://arxiv.org/abs/1012.4986}{{\tt arXiv:1012.4986}}.

\bibitem{Nakayama:2016cim}
Y.~Nakayama, \emph{{Bootstrapping critical Ising model on three-dimensional
  real projective space}},
  \href{http://dx.doi.org/10.1103/PhysRevLett.116.141602}{Phys. Rev. Lett. {\bf
  116} (2016)  141602},
\href{http://arxiv.org/abs/1601.06851}{{\tt arXiv:1601.06851 [hep-th]}}.

\bibitem{Quella:2006de}
  T.~Quella, I.~Runkel and G.~M.~T.~Watts,
  \emph{Reflection and transmission for conformal defects},
  JHEP {\bf 0704} (2007) 095,
 \href{http://dx. doi:10.1088/1126-6708/2007/04/095}
  { \tt arXiv:hep-th/0611296}.

\bibitem{Billo:2016cpy}
  M.~Billò, V.~Goncalves, E.~Lauria and M.~Meineri,
  \emph{Defects in conformal field theory},
  JHEP {\bf 1604} (2016) 091
  doi:10.1007/JHEP04(2016)091
  {\tt arXiv:1601.02883 [hep-th]}.
  
\bibitem{Abe:1981} 
   R.~Abe,
  \emph{d' Dimensional Defect in d-Dimensional Lattice. I 
 Nonuniversal Local Critical  Exponent in the Limit $n\to \infty$},
 Prog.\ Theor.\ Phys.\  {\bf 65}, 1237 (1981).
  

\bibitem{McCoy:1980ag}
B.~M. McCoy and J.~H.~H. Perk, \emph{{Two Spin Correlation Functions of an
  Ising Model With Continuous Exponents}},
\href{http://dx.doi.org/10.1103/PhysRevLett.44.840}{Phys. Rev. Lett. {\bf 44}
  (1980)  840}.

\bibitem{Oshikawa:1996dj}
M.~Oshikawa and I.~Affleck, \emph{{Boundary conformal field theory approach to
  the critical two-dimensional Ising model with a defect line}},
  \href{http://dx.doi.org/10.1016/S0550-3213(97)00219-8}{Nucl. Phys. {\bf B495}
  (1997)  533--582},
\href{http://arxiv.org/abs/cond-mat/9612187}{{\tt arXiv:cond-mat/9612187
  [cond-mat]}}.

\bibitem{Hasenbusch:2011s}
M.~Hasenbusch, \href{http://dx.doi.org/10.1103/PhysRevB.84.134405}{\emph{{Monte
  Carlo study of surface critical phenomena: The special point}},Phys. Rev.
  {\bf B 84} ({Oct}, {2011})  134405},
  \href{http://arxiv.org/abs/1108.2425}{{\tt arXiv:1108.2425}}.
\url{{http://link.aps.org/doi/10.1103/PhysRevB.84.134405}}

\bibitem{Dolan:2003hv}
F.~Dolan and H.~Osborn, \emph{{Conformal partial waves and the operator product
  expansion}},
  \href{http://dx.doi.org/10.1016/j.nuclphysb.2003.11.016}{Nucl.Phys. {\bf
  B678} (2004)  491--507},
\href{http://arxiv.org/abs/hep-th/0309180}{{\tt arXiv:hep-th/0309180 [hep-th]}}.

\bibitem{Weinberg:2012cd}
S.~Weinberg, \emph{{Minimal fields of canonical dimensionality are free}},
  \href{http://dx.doi.org/10.1103/PhysRevD.86.105015}{Phys. Rev. {\bf D86}
  (2012)  105015},
\href{http://arxiv.org/abs/1210.3864}{{\tt arXiv:1210.3864 [hep-th]}}.

\bibitem{McAvity:1995zd}
D.~McAvity and H.~Osborn, \emph{{Conformal field theories near a boundary in
  general dimensions}},
  \href{http://dx.doi.org/10.1016/0550-3213(95)00476-9}{Nucl.Phys. {\bf B455}
  (1995)  522--576},
\href{http://arxiv.org/abs/cond-mat/9505127}{{\tt arXiv:cond-mat/9505127
  [cond-mat]}}.

\bibitem{Diehl:1998mh}
H.~Diehl and M.~Shpot, \emph{{Massive field theory approach to surface critical
  behavior in three-dimensional systems}},
  \href{http://dx.doi.org/10.1016/S0550-3213(98)00489-1}{Nucl.Phys. {\bf B528}
  (1998)  595--647},
\href{http://arxiv.org/abs/cond-mat/9804083}{{\tt arXiv:cond-mat/9804083
  [cond-mat]}}.

\bibitem{Diehl:1981jgg}
H.~W. Diehl and S.~Dietrich, \emph{{Field-theoretical approach to static
  critical phenomena in semi-infinite systems}},
\href{http://dx.doi.org/10.1007/BF01298293}{Zeit. Phys. {\bf B42} (1981)
  65--86}.

\bibitem{Diehl:1981}
H.~Diehl and S.~Dietrich, \emph{{Field-theoretical approach to multicritical
  behavior near free surfaces}},
\href{http://dx.doi.org/10.1103/PhysRevB.24.2878}{Phys.Rev. {\bf B24} (1981)
  2878--2880}.

\bibitem{Diehl:1983}
H.~Diehl and S.~Dietrich, \emph{{Multicritical behaviour at surfaces}},
Zeit. f. Phys B {\bf 50} (1983)  117.

\bibitem{Deng:2005dh}
Y.~J. Deng, H.~W.~J. Bl{\"o}te, and M.~P. Nightingale, \emph{{Surface and bulk
  transitions in three-dimensional O(n) models}},
  \href{http://dx.doi.org/10.1103/PhysRevE.72.016128}{Phys. Rev. {\bf E72}
  (2005)  016128--016138},
\href{http://arxiv.org/abs/cond-mat/0504173}{{\tt arXiv:cond-mat/0504173}}.

\bibitem{Hasenbusch:2011yya}
M.~Hasenbusch, \emph{{Finite size scaling study of lattice models in the
  three-dimensional Ising universality class}},
  \href{http://dx.doi.org/10.1103/PhysRevB.82.174433}{Phys.Rev. {\bf B82}
  (2010)  174433},
\href{http://arxiv.org/abs/1004.4486}{{\tt arXiv:1004.4486}}.

\end{thebibliography}
\end{document}